\documentclass[12pt]{iopart}

\usepackage[caption=false]{subfig}

\usepackage[T1]{fontenc}
\usepackage[utf8]{inputenc}
\usepackage[english]{babel}

\usepackage{graphicx}
\usepackage{enumitem}
\usepackage{textcomp}
\usepackage{gensymb}
\usepackage{siunitx}
\usepackage{isotope}
\usepackage{rotating}

\usepackage[newfloat,frozencache,cachedir=.]{minted}
\setminted{fontsize=\footnotesize}
\usepackage{caption}
\newenvironment{code}{\captionsetup{type=listing}}{}
\SetupFloatingEnvironment{listing}{name=Source code}

\usepackage[sort&compress]{natbib}
\expandafter\let\csname equation*\endcsname=\relax
\expandafter\let\csname endequation*\endcsname=\relax
\usepackage{amsmath}

\usepackage{amsfonts,amssymb,amscd}
\usepackage{xcolor, soul}
\usepackage{xspace}
\usepackage{microtype}

\usepackage{hyperref}
\usepackage{cleveref} 

\usepackage{fdsymbol}

\sethlcolor{yellow}

\hypersetup{colorlinks, linkcolor={red!50!black}, citecolor={blue!50!black}, urlcolor={blue!80!black}}
\microtypesetup{
	protrusion=alltext-nott,
	expansion=alltext-nott,
	final
}

\graphicspath{{figures/}}

\begin{document}
\title[GATE 10 - Part II]{GATE 10 Monte Carlo particle transport simulation - Part II: architecture and innovations}

\author{
Nils Krah\textsuperscript{1,15},
Nicolas Arbor\textsuperscript{9},
Thomas Baudier\textsuperscript{1},
Julien Bert\textsuperscript{4},
Konstantinos Chatzipapas\textsuperscript{4,16},
Martina Favaretto\textsuperscript{5},
Hermann Fuchs\textsuperscript{10},
Loïc Grevillot\textsuperscript{5},
Hussein Harb\textsuperscript{4},
Gert Van Hoey\textsuperscript{17},
Maxime Jacquet\textsuperscript{1},
Sébastien Jan\textsuperscript{6},
Yihan Jia\textsuperscript{5},
George C. Kagadis\textsuperscript{12},
Han Gyu Kang\textsuperscript{3},
Paul Klever\textsuperscript{8,13},
Olga Kochebina\textsuperscript{6},
Wojciech Krzemien\textsuperscript{18},
Lydia Maigne\textsuperscript{7},
Philipp Mohr\textsuperscript{13},
Guneet Mummaneni\textsuperscript{2},
Valentina Paneta\textsuperscript{11},
Panagiotis Papadimitroulas\textsuperscript{11},
Alexis Pereda\textsuperscript{7},
Axel Rannou\textsuperscript{4},
Andreas F. Resch\textsuperscript{5},
Emilie Roncali\textsuperscript{2},
Maxime Toussaint\textsuperscript{14},
Carlotta Trigila\textsuperscript{2},
Charalampos Tsoumpas\textsuperscript{13},
Jing Zhang\textsuperscript{4},
Karl Ziemons\textsuperscript{8},
David Sarrut\textsuperscript{1}
}

\address{
\textsuperscript{1} Université de Lyon; CREATIS; CNRS UMR5220; Inserm U1294; INSA-Lyon; Université Lyon 1, Lyon, France.\\
\textsuperscript{2} University of California, Davis, Davis CA USA\\
\textsuperscript{3} National Institutes for Quantum Science and Technology (QST), 4-9-1 Anagawa, Inage-ku, Chiba, Japan\\
\textsuperscript{4} LaTIM, INSERM UMR1101, University of Brest, Brest, France\\
\textsuperscript{5} MedAustron Ion Therapy Center, Wiener Neustadt, Austria\\
\textsuperscript{6} Université Paris-Saclay, Inserm, CNRS, CEA, Laboratoire d’Imagerie Biomédicale Multimodale (BioMaps), Orsay, France\\
\textsuperscript{7} Université Clermont Auvergne, Laboratoire de Physique de Clermont Auvergne, CNRS, Clermont-Ferrand, France.\\
\textsuperscript{8} FH Aachen University of Applied Sciences, Germany\\
\textsuperscript{9} Université de Strasbourg, IPHC, CNRS, UMR7178, F-67037 Strasbourg, France.\\
\textsuperscript{10} Medical University of Vienna, Vienna, Austria\\
\textsuperscript{11} Bioemission Technology Solutions, BIOEMTECH, Athens, Greece\\
\textsuperscript{12} 3DMI Research Group, Department of Medical Physics, University of Patras, Rion, Greece\\
\textsuperscript{13} University of Groningen, University Medical Center Groningen, Groningen, Netherlands\\
\textsuperscript{14} Laboratoire CRCI2NA, INSERM, CNRS, Nantes Université, Nantes, France\\
\textsuperscript{15} Department of Research and Development, Holland Proton Therapy Centre Delft, Delft, The Netherlands\\
\textsuperscript{16} Department of Radiation Science and Technology, Technical University of Delft, Delft, The Netherlands\\
\textsuperscript{17} XEOS, Ghent, Belgium\\
\textsuperscript{18} High Energy Physics Division, National Centre for Nuclear Research, Andrzeja Soltana 7, Otwock, Swierk, PL-05-400, Poland\\
}

\newpage

\begin{abstract}
Over the past years, we have developed GATE version 10, a major re-implementation of the long-standing Geant4-based Monte Carlo application for particle and radiation transport simulation in medical physics. This release introduces many new features and significant improvements, most notably a Python-based user interface replacing the legacy static input files. The new functionality of GATE version 10 is described in the part~1 companion paper~\citep{Sarrut2025}. 

The development brought significant challenges. In this paper, we present the solutions that we have developed to overcome these challenges. In particular, we present a modular design that robustly manages the core components of a simulation: particle sources, geometry, physics processes, and data acquisition. The architecture consists of parts written in C++ and Python, which needed to be coupled. We explain how this framework allows for the precise, time-aware generation of primary particles, a critical requirement for accurately modeling positron emission tomography (PET), radionuclide therapies, or prompt-gamma timing systems. We present how GATE~10 handles complex Geant4 physics settings while exposing a simple interface to the user. Furthermore, we describe the methodological solutions that facilitate the seamless integration of advanced physics models and variance reduction techniques. The architecture supports sophisticated scoring of physical quantities (such as Linear Energy Transfer and Relative Biological Effectiveness) and is designed for multithreaded execution. The new user interface allows researchers to script complex simulation workflows and directly couple external tools, such as artificial intelligence models for source generation or detector response. By detailing these architectural innovations, we demonstrate how GATE 10 provides a more powerful and flexible tool for research and innovation in medical physics.

This paper is not intended to be a developer guide. Its purpose is to share with the research community in-depth explanations of our development effort that made the new GATE~10 possible. 

       
\end{abstract}



\section{Introduction}
\label{sec:intro}



GATE 10 is a completely new version of the long-standing Monte Carlo application GATE, built upon the Geant4 toolkit and designed for medical physics and medical imaging. This new release introduces a more powerful and flexible paradigm for simulating complex physical phenomena. Compared to previous versions and other simulation toolkits, the user now sets up and runs simulations using Python instead of static input files. This aspect not only simplifies the simulation set up, but also opens the door to functionalities not previously available, or only indirectly. The user can make full use of Python and third-party libraries to configure components and input to the simulation directly within the same script that runs the simulation. Other new features include the possibility to run a simulation multi-threaded or in (multiple) sub-processes. While the companion paper \citep{Sarrut2025} provides a description of these new features and functionalities, this article (Part 2) focuses on the underlying architectural innovations and the technical solutions developed to realize this new version. 


The development of this new framework required solutions to several challenges, particularly in creating a hybrid architecture that combines the computational speed of C++ for particle transport with the flexibility of Python for simulation control. Moreover, it must comply with some limitations imposed by the Geant4 toolkit~\citep{allison2016}, e.g. two simulations cannot be run in the same process. 
The primary goal was to harness the power of the Geant4 toolkit within an accessible environment, allowing physicists to focus on the application rather than on complex programming. 

This paper aims to highlight these foundational innovations and to demonstrate how they provide a more powerful and versatile tool for the medical physics community.

\section{Simulation architecture}
\label{sec:architecture}


In this section, we explain the architecture of GATE~10 and how the user interface, i.e. the simulation script, links to GATE's internal mechanisms. 
GATE~10's architecture is the result of technical requirements (e.g., interaction Python/C++) and a series of design considerations. 
GATE users come from a variety of backgrounds with mixed experience in software engineering and computer science.
They often have a reasonable understanding of Monte Carlo particle transport simulations and possibly some background with Geant4. Implementing a Monte Carlo simulation should not require a high level of programming skills, yet, at the same time, GATE should allow users to extensively exploit the functionalities of Python and third-party Python packages, for data processing, statistical analysis, visualization, or physics databases for example.
Geant4 functionalities relevant for simulations in medical physics and medical imaging should be accessible without requiring a deep understanding of the Geant4 toolkit. 


The users should be able to configure and run a GATE simulation from a single Python script, and, if desired, they should be able to access and post-process the simulation output, e.g. dose maps or phase space distributions, from the same script as well. It should be possible to run a simulation multiple times from the same script to vary and test certain parameters. A simple simulation, e.g. standard geometry and physics, should require only a simple, short Python script to configure it. Parameters should always have a default value and components should have a default behavior so that the user only has to configure those that really require customization in a given simulation. At the same time, it should be possible to implement a more complex system such as a PET scanner with non-standard geometries and provide it via a module to other researchers who then use it in their script to simulate, for example, the PET acquisition of a particular subject. The order in which parameters of a GATE simulation are defined in a configuration script should be irrelevant, and it should be possible to change several parameters among multiple executions of the same simulation using a single script, e.g. with a for-loop. 


To make it easier to follow the remainder of this paper, we provide a short example of a simulation script (source code~\ref{code:example-script}) and refer to it in what follows. 

\begin{code}
\caption{Example simulation script}
\label{code:example-script}
\begin{minted}[xleftmargin=20pt,linenos]{python}
import opengate as gate
sim = gate.Simulation()
mm = gate.g4_units.mm
MeV = gate.g4_units.MeV
patient = sim.add_volume("Image", name="patient")
patient.image = "patient-4mm.mhd"
patient.mother = "world"
patient.voxel_materials = [
    [-2000, -900, "G4_AIR"],
    [-900, -100, "Lung"],
    [-100, 0, "G4_ADIPOSE_TISSUE_ICRP"],
    [0, 300, "G4_TISSUE_SOFT_ICRP"],
    [300, 800, "G4_B-100_BONE"],
    [800, 6000, "G4_BONE_COMPACT_ICRU"]]
patient.set_production_cut(
    particle_name="electron",
    value=3 * mm)
src = sim.add_source("GenericSource", name="p_src")
src.particle = "proton"
src.position.type = "sphere"
src.position.radius = 10 * mm
src.position.translation = [0, 0, -140 * mm]
src.n = 1e4 # number of primary particles
src.direction.type = "momentum"
src.direction.momentum = [0, 0, 1]
dose_actor = sim.add_actor("DoseActor", "dose")
dose_actor.attached_to = "patient"
dose_actor.size = [99, 99, 99]
dose_actor.spacing = [2 * mm, 2 * mm, 2 * mm]
dose_actor.output_coordinate_system = 'attached_to_image'
dose_actor.translation = [2 * mm, 3 * mm, -2 * mm]
for en in [130, 140, 150, 160]: 
    src.energy.mono = en * MeV
    dose.output_filename = f"patient_energy_{en}.mhd"
    sim.run(start_new_process=True)
\end{minted}
\end{code}

\subsection{Combining C++/Geant4 particle tracking with Python a user interface}

\begin{figure}[htbp]
  \centering
  \includegraphics[width=0.9\columnwidth]{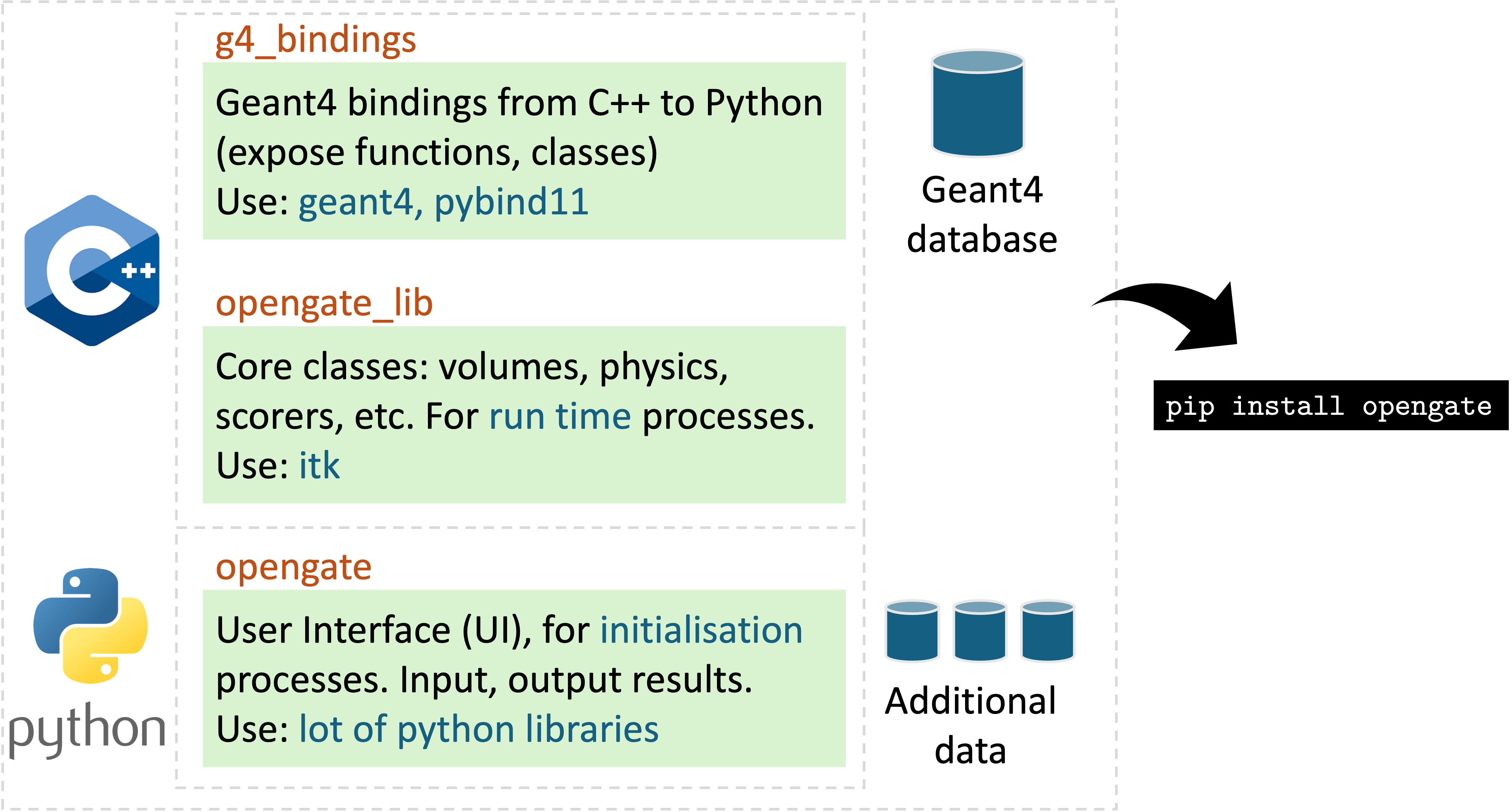}
  \caption{The main components of the unique 'opengate' package: Geant4 bindings, C++ runtime library, Python interface library and associated database. Everything is installed by a single \texttt{pip install} command.}
  \label{fig:overall_schema}
\end{figure}

GATE~10 leverages two programming languages at the same time, namely Python and C++: the former because it is flexible and easy to use and the latter because it results in optimized, computationally efficient, compiled code. 
In the following, we describe the architecture that we developed to make this challenging synergy work.  
Specifically, GATE~10 consists of a high-level Python package and two C++ libraries (see figure~\ref{fig:overall_schema}). 

The first C++ library is a low-level Python wrapper for selected parts of Geant4, named \verb|g4_bindings|. It directly exposes Geant4 classes and functions, which can then be called from Python. It is similar to the \verb|g4py| and \verb|g4python|\footnote{\url{https://github.com/koichi-murakami/g4python}} packages that were proposed by the Geant4 collaboration in the past to steer Geant4 simulations via Python. 

The second C++ component, \verb|opengate_lib|, contains several classes (around 100)
 and implements performance critical functionality of GATE~10. It is an evolution of the historic core of GATE that has been described in previous GATE publications (see \citet{sarrut2022} for the latest reference). The classes in \verb|opengate_lib| manage key simulation elements that deeply hook into the nested Geant4 event loop, e.g. to generate new primary particles (sources, timing) or to score information about the simulation (actors), e.g. the dose deposited in a volume or the properties of particles passing a detector surface.

Both C++ libraries are bound to Python with pybind11\footnote{\url{https://github.com/pybind/pybind11}}. 
This binding is essential because the Python side and the C++ side need to communicate with each other. 
A Python function must be able to call a function implemented in C++, pass variables to it, and access the returned value. Another example is that a Python function needs to be able to create a new Geant4 object by calling the constructor of a C++ class, e.g. to create a new instance of \verb|G4Sphere| when creating a spherical volume. In the opposite direction, many C++-implemented classes might need access to a Python dictionary where user input parameters are stored, and some even need to call a Python function to trigger some action or to access a returned value. 

To achieve this, we implemented a pybind11 configuration for each C++ class to be wrapped. Specifically, such a file defines by which name the C++ class is visible to Python and which member functions are exposed. 
Inheritance required special attention, necessitating the use of an intermediary wrapper class, sometimes referred to as \emph{trampolines}, to enable overriding virtual functions in Python.

\subsection{The main Python user package} 

The third part of GATE~10 is the main Python package named \verb|opengate|. It manages the overall simulation architecture, sets up the Geant4 components of the simulation, and handles user-defined parameters. 
The underlying concept is to separate the user interaction (Python) from the actual simulation (C++). For example, the plausibility of user parameters is checked by the Python package before triggering a Geant4 run. 
Separating user interaction and core functionality in this way also allows for automatic default parameter values and automatic generation of associated documentation. 

The architectural concept is to divide the configuration and the execution aspects of a simulation.  
All components of a GATE~10 simulation, e.g. actors, volumes, sources, are implemented as Python classes that inherit from a common base class called \verb|GateObject|. This base class and an associated factory mechanism handle most of the configuration by the user, as we will explain in section~\ref{sec:gateobject}. Each component is associated with a manager, e.g. the \verb|VolumeManager|, \verb|ActorManager|, \verb|SourceManager|, etc., which will be explained in section~\ref{sec:managers}. Execution a simulation is handled via a set of \verb|engines|, presented in section~\ref{sec:engines}. 


\subsection{Main base class: GateObject}
\label{sec:gateobject}

A large part of the functionality in GATE~10 relies on mechanisms implemented in a common Python base class called \verb|GateObject|. 
Every class inheriting from \verb|GateObject| is enhanced via a class factory mechanism that is triggered automatically in two ways: (1) by explicit commands inside the modules that are called whenever \verb|opengate| is imported, and (2) via the \verb|__new__()| method whenever an instance of the respective class is created. The latter mechanism is important, for example, when a simulation is run within a sub-process. This factory mechanism remains invisible to the user and greatly improves maintainability and consistency. 

Almost every component of a GATE~10 simulation, represented as a Python object, needs to handle user input parameters. It was a fundamental design choice to encapsulate this information in a dedicated structure to separate user parameters from other attributes of the Python class. 
At the same time, we wanted the user to access each parameter with the shortest possible syntax. 
We achieved this by automatically creating setter and getter properties for each user input parameter. 
These setter and getter hooks can also trigger certain actions whenever a user sets a user parameter, e.g.
to ensure that other dependent options are active as well, to perform type checking, or to convert the input, e.g. from a list to a Numpy array, etc. For example, whenever the parameter ``mother'' of a volume is changed, the associated setter hook informs the volume manager that it needs to update the volume hierarchy tree.  



Through the class factory (see above), we implemented a mechanism to detect whenever a user attempts to set an unknown user input parameter. This is important because Python allows for dynamic attribute definition and the user would remain unaware, e.g. of a mistyped parameter name. 



Every GATE~10 object based on the GateObject class can be serialized, i.e. converted into a basic representation, in two alternative ways: first, to be passed on to a subprocess (see section~\ref{sec:subprocess}) and second, to be stored in a human-readable text-based format. This is possible thanks to dedicated encoders and the fact that all user parameters are structurally kept apart from the remaining attributes. The ``backwards'' direction is also possible: a new GATE~10 object, e.g. a certain volume, can be created starting from a serialized representation of the object. This is the crucial element underlying the mechanism by which GATE~10 simulations can be saved and loaded as JSON files (see section~3.4 in the part 1 companion paper). We also think that this built-in functionality will be a cornerstone in future mechanisms to dispatch GATE~10 simulations to computation clusters.

\subsection{Managers}
\label{sec:managers}


GATE~10 handles the components of a simulation via a series of managers, each implemented as a class inheriting from the common base class \verb|GateObject| (see section~\ref{sec:gateobject}). The main manager class is called 
\verb|Simulation| and all GateObjects refer to it. The \verb|Simulation| object delegates certain responsibilities to specific managers, which include the \verb|VolumeManager|, \verb|PhysicsManager|, \verb|ActorManager|, \verb|SourceManager|. An important task of all managers is to keep an inventory of all the Python objects that represent the components of the simulation for which the manager is responsible. 
For example, the \verb|attached_to| parameter of an actor is a string referring to a volume and the actor can retrieve the actual Python object via the \verb|VolumeManager|. 
All managers implement an \verb|add| function, e.g. \verb|add_actor|, \verb|add_volume|, which performs consistency checks before inserting an object into the inventory dictionary.

In addition to bookkeeping, the managers fulfill specific tasks. For example, the \verb|VolumeManager| handles the nesting hierarchy of volumes and keeps it up to date. Whenever a new volume is added to the simulation with \verb|add_volume()|, the volume manager triggers an update of the geometry hierarchy. This is also done via a setter hook whenever the user sets (or re-sets) the parameter \verb|mother| of any volume already known to the simulation. The hierarchy itself is implemented through the external Python package \verb|anytree| leveraging a special attribute \verb|parent|. Thanks to this architecture, GATE~10 can loop over all volumes in a hierarchical fashion and visualize the geometry hierarchy without depending on Geant4. This is useful for extracting information about parts of the volume hierarchy, e.g. representing the crystals in a PET system. 

The \verb|PhysicsManager| handles production cuts, user limits such as step limiters, and special physics such as optical physics. Production cuts, i.e. thresholds that determine under which conditions Geant4 should create and track secondary particles, are a representative example where GATE~10 leverages a complex internal mechanism while exposing a simple interface to the user. We explain this in more detail in section~\ref{sec:regions_cuts}.

\subsection{Engines}
\label{sec:engines}

\begin{figure}[htpb]
  \centering
  \includegraphics[width=0.8\columnwidth]{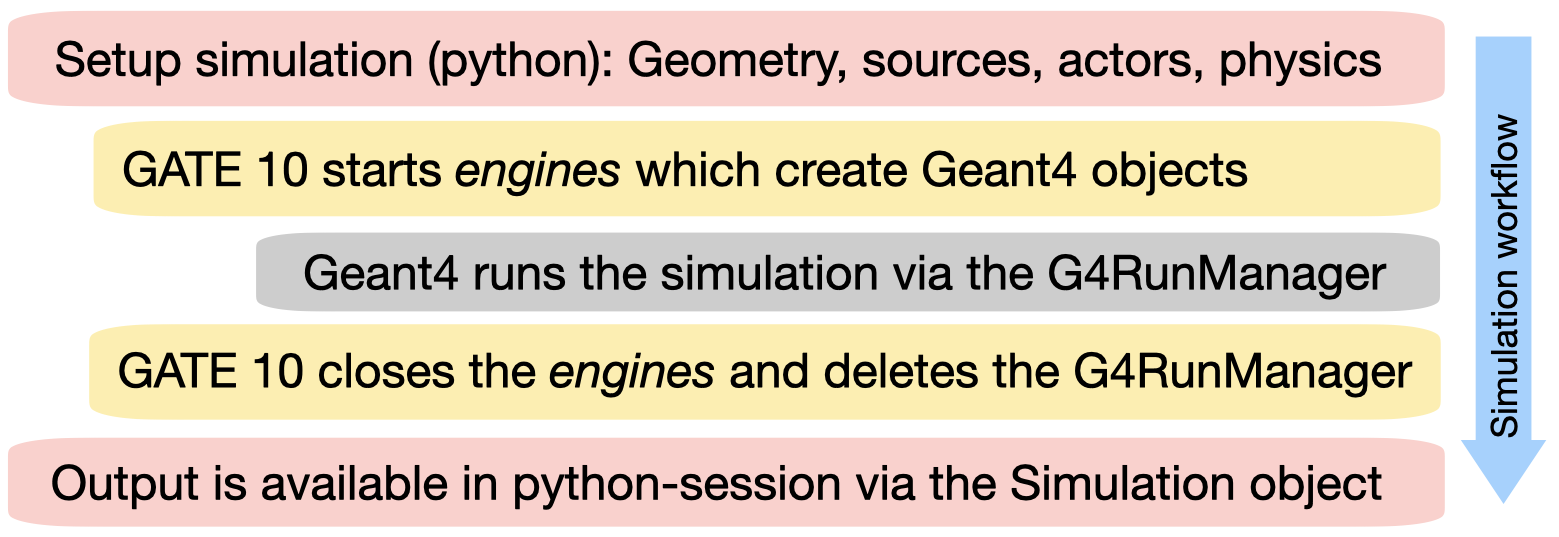}
  \caption{Workflow of a GATE~10 simulation from set-up to simulation run.}
  \label{fig:workflow}
\end{figure}

GATE~10 implements a series of \verb|Engines|, i.e. Python classes, whose responsibility is to initialize and run the Geant4 simulation. The engines are organized in a similar fashion as the managers, with the \verb|SimulationEngine| being the main one complemented by an engine for each logical part of the simulation, namely a \verb|PhysicsEngine|, \verb|VolumeEngine|, \verb|ActorEngine|, \verb|SourceEngine|,and \verb|ActionEngine|. 
The general concept is shown in figure~\ref{fig:workflow}: When the user executes a simulation with \verb|sim.run()| (see example~\ref{code:example-script}), GATE~10 creates a set of engines which in turn handle the creation of all Geant4 objects, initialize the Geant4 simulation and trigger the generation of primary particles (source). Output data is stored on disk and/or kept in memory, depending on the actor. When the Geant4 simulation ends, all previously created Geant4 objects are deleted and/or de-referenced (see section~\ref{sec:life-cycle-management}), and the engines are closed. Only the pure Python part of the simulation remains, namely the managers and the Python representation of the simulation components. In this way, the user can continue with post-processing steps directly in the same Python script and, depending on the actor, directly access output. In example~\ref{code:example-script}, the user could access the deposited energy map via \verb|dose_actor.edep.image| as an ITK image, or convert it to a Numpy array via \verb|np.asarray(dose_actor.edep.image)|. 

By decoupling the configuration and the execution of a simulation, GATE~10 is able to execute a simulation multiple times within the same Python script. This is a drastic enhancement over previous versions of GATE and over Geant4 itself. In example~\ref{code:example-script}, the simulation is executed four times in a "for-loop", each time with a different beam energy, but any other simulation parameter could be changed as well. 
During each iteration, the \verb|sim.run()| command triggers the creation of a new set of engines and GATE~10 goes through the entire workflow in figure~\ref{fig:workflow}. 
Such a repeated execution of a simulation from the same script is enabled by the engine architecture, with GATE~10 transferring the simulation to a new sub-process each time, as explained in section~\ref{sec:subprocess}. It should be highlighted that Geant4 itself supports multiple runs within the same Geant4 instance, which is useful when simulating dynamic geometries such as a living subject or rotating gantry, for example. GATE~10 uses the Geant4 run mechanism with the concept of dynamic parameterisation which we describe in the part~1 companion paper.

It is essential to note that GATE sets up the simulation with the Geant4 run manager, so it behaves like a Geant4 simulation. 
Specifically, the method \verb|SimulationEngine.run_engine()| is similar to the main executable of a Geant4 simulation because it first initializes the simulation and then turns on the primary particles generator. 
The initialization in GATE~10 closely follows the \verb|G4RunManager|'s initialization sequence, but interleaves it with GATE~10 specific initialization steps. To this end, GATE~10 uses a slightly modified version of the \verb|G4RunManager| wrapped via pybind11. 


Following the object-oriented nature of Python, initialization is delegated to the respective object. 
For example, the \verb|VolumeEngine.Construct()| method loops through the volume hierarchy, via the \verb|VolumeManager|, and calls the \verb|construct()| method implemented in each volume class. This latter method creates all necessary Geant4 objects such as the Solid, \verb|G4LogicalVolume|, \verb|G4PVPlacement|, as well as material and other specific Geant4 objects. 
This architecture keeps the code structured and simplifies maintenance.


\section{Specific technical challenges}\label{sec:technical-challenges}

In addition to the general challenge of coupling the C++ layer of Geant4 with the Python layer, which has been addressed via the architecture presented in the previous section~\ref{sec:architecture}, we describe other novel technical aspects  in GATE 10 below.

\subsection{Life cycle and resource management between Python and C++}\label{sec:life-cycle-management}

Python and C++ both use the concept of references to objects in memory, but differ in life cycle management, i.e. how they delete objects from memory. 
It is imperative to implement this correctly, otherwise two main issues arise. First, an object might remain in memory needlessly, which progressively reduces the computer's accessible memory. Second, if an object is deleted from memory, but the program still has a reference to it, a segmentation fault will occur, i.e. the simulation will crash. Life cycle management was particularly challenging not only because GATE~10 is a hybrid Python/C++ software, but also because Geant4 has its own rather complex life cycle management that must be considered. 

One important issue arose from Geant4 objects created from the Python part of GATE~10 via a call to the constructor of the C++ class. Consequently, a Python reference to a C++ object is stored in the memory. For example, every \verb|Volume| object holds a reference to a \verb|G4LogicalVolume| once the simulation is initialized. At the end of a Geant4 simulation, the \verb|G4RunManager|, by Geant4 design, deletes all user-created objects, including e.g. \verb|G4LogicalVolume|. 
At this point, a segmentation fault could occur if the Python-side of GATE~10 still holds a reference to the deleted object. To avoid this, we implemented the following protocol: Every \verb|GateObject| Python class has a \verb|close()| method which is called when the engines are shut down at the end of the simulation. The \verb|close()| method, among other actions, resets references to Geant4 objects. 
This mechanism guarantees that no references to Geant4 objects remain when a simulation is terminated (either prematurely aborted or completed) and effectively prevents segmentation faults. The mechanism is invisible to the user and easy to implement for the developer who only needs to ensure that the close() method resets all references. 

The interplay between life-cycle management in Python and C++ also required care from the C++ perspective. In particular, Python's garbage collection mechanism automatically deletes an object from memory when no reference to it exists anymore (or at any later stage). However, if the C++ side still has a pointer to the object and tries to delete the object with its destructor, a segmentation fault will occur. To avoid this, we adapted the \verb|pybind| wrapping (\emph{nodelete} keyword) to instruct Python not to delete the respective object. 

\subsection{Callbacks from Geant4 to Python}\label{sec:callbacks}

In previous sections, we discussed how the configuration and initialization of a simulation are primarily handled from the interpreted Python side, while particle tracking and scoring are performed by the C++ compiled library once the simulation has started.  
Specifically, the Geant4 kernel implements a hierarchy of loops (primary particles, tracks, steps, etc.) within which certain functions are called, e.g. \verb|BeginOfEventAction| or \verb|SteppingAction|. 
Most of these functions are implemented in C++, i.e. they are either part of the Geant4 toolkit or the GATE core library (mainly its actors and sources). It is, however, possible, to implement any such function in Python and call it from the C++ kernel. We refer to this as a \emph{callback}. Examples are the \verb|EndOfRunAction| functions of certain actors that handle the file output in Python. It is also possible to mix both approaches, i.e. implement a partial function in Python while calling another part from C++. 

Callbacks from the Geant4/C++ side to Python functions are particularly useful for advanced simulation setups where specific actions need to be triggered based on the simulation state or events. A typical example is triggering from C++ the execution of a neural network in Python at regular intervals, such as after simulating a batch of particles, as seen in ARF~\citep{sarrut2018} or GAN sources~\citep{sarrut2021}. Callbacks can be used to modify simulation parameters on-the-fly, log detailed information, or integrate it with external systems for data processing. To enable these callbacks, we utilized a so-called trampoline pattern in pybind11, which serves as a bridge between C++ and Python. This offers the advantage of Python classes inheritance from C++ classes. It allows Python classes to override virtual methods defined in C++ base classes and seamlessly integrate C++ and Python portions of the code. 

While callbacks are useful, they introduce a time cost due to the synchronization and type conversion overhead of transitioning from compiled, statically typed, optimized C++ code into a dynamically typed, interpreted Python environment. 
Our experience so far has shown that this overhead is negligible for functions which are called only infrequently during a simulation, e.g. once per Geant4 run, but have a noticeable impact when functions are called at every new track or step. If a specific application requires a callback for these latter functions, e.g. to invoke a neural network at each step via a Python library, it is recommended to use a buffer mechanism that triggers the callbacks in batches, for example, every 100,000 events. This approach ensures that the overhead remains negligible compared to other parts of the simulation, maintaining overall performance. 

The following example illustrates the effective use of callbacks in GATE version 10: A GATE actor accumulates particles that impinge on a detector plane in a SPECT simulation. Once the buffer reaches its capacity, a callback to the Python side is triggered. The Python function processes these particles as input to a neural network model, in this case modeling an angular response function (ARF), updating the image accordingly. After the callback completes, the buffer is reset to zero and particle tracking in C++ resumes. 


\subsection{Running multiple simulations}\label{sec:subprocess}

The Geant4 engine is designed with the constraint that it can only be executed once within a single process. This limitation arises from its complex architecture which includes multiple singleton pointers and intricate memory allocation and de-allocation mechanisms. 
This would limit the versatility of GATE~10 because users could execute a simulation only once within a single GATE~10 Python script, e.g. a "for-loop" as shown in example~\ref{code:example-script} would not be possible. Likewise, the user could not run a simulation multiple times in an interactive Python interpreter such as \verb|jupyter notebook| or \verb|ipython|. 
Multiple executions, however, are very convenient for example, to explore a series of simulation parameters or to estimate statistical uncertainty. 


To address this issue, we developed a mechanism based on the engine architecture described in section~\ref{sec:engines} through which the user can run simulations in a separate process. This capability is easily accessible to users through the keyword argument \verb|start_new_process=True| in the \verb|run()| command (line~35 in example~\ref{code:example-script}).  
When this option is enabled, the main manager, implemented in the Python class \verb|Simulation| (see section~\ref{sec:managers}),  dispatches the creation of all necessary engines to a new subprocess. The engines then set up and run the Geant4 engine in the subprocess. When the Geant4 simulation ends, it is destroyed, the engines are closed, and the output is returned to the main process in which the user script is running. 
The underlying mechanism remains invisible to the user. 

Making GATE~10 dispatchable to a subprocess required careful architecture design and involved several complex steps. 
We relied on the \verb|multiprocessing| package from Python's standard library and specifically on the method ``spawn'' to start a new process. 
When executing \verb|sim.run(start_new_process=True)|, the multiprocessing package starts a fresh Python process within which all Python objects need to be recreated to reflect the same state at which they were before. This is achieved via serialization handled by the built-in \verb|pickle| module: a Python object is encoded in a chain of bytes, sent to the subprocess, and decoded again. Implementing this backbone mechanism has been challenging in GATE~10 because of its hybrid C++/Python nature. For example, many C++ implemented objects (classes) are not serializable and require special handling. Furthermore, the pickling mechanism must be made aware of any instantiation and initialization steps implemented in C++, e.g. in the constructor. Much of this mechanism is implemented in the \verb|GateObject| base class (see section~\ref{sec:gateobject}) and automatically available in any class inheriting from it. Additionally, specific classes implement their own mechanisms to be serialized and restored correctly. 


\subsection{Consistent handling of scorer output}
\label{sec:actoroutput}

Actors are GATE's way of scoring information from a simulation, e.g. the energy deposited by particles in a medium or the properties of particles entering a certain volume. Technically speaking, actors are classes that implement certain predefined functions that are called by Geant4 during the iterative particle transport, e.g. when a new primary particle is generated or at every step taken by the simulation. Some actors merely modify the course of the simulation, such as the KillActor that removes particles from the simulations under certain conditions, but most actors actually generate output data. The \verb|DoseActor| (and variants of it), for example, generate three-dimensional dose maps, i.e. on a regular grid. Actors that register particle-related information, such as the \verb|PhaseSpaceActor|, typically generate list-mode data where each list entry is a set of properties, e.g. position, direction, energy. 

In versions of GATE prior to the new GATE~10, the handling of output data was implemented \emph{ad-hoc} by each individual actor, usually by writing data to disk at a certain point during the simulation. In GATE~10, we have developed an abstract layer, i.e. a set of classes, which is responsible for handling actor output and which every actor uses. The actor output classes are responsible for consistently handling the file location with respect to the simulation's output directory, to hold output data in memory if requested by the user (and if technically possible), and to keep track of runtime intervals. The latter is important in practice because it makes it possible to keep and/or store output data associated with each run of a simulation individually. An example would be the dose maps associated with breathing phases in a radiotherapy simulation. 

The actor output layer and associated classes are also able to automatically merge output coming from distinct simulation runs. Furthermore, the actor output layer allows GATE~10 to easily recover output from another process when the simulation is run in a sub-process (see section~\ref{sec:subprocess}). We anticipate that this mechanism will play a vital role in the future when developing automatic mechanisms to parallelize simulations.  

Finally, the actor output layer makes code maintenance significantly easier because actors share a common internal infrastructure. It also reduces code redundancy and keeps functionality consistent across actors.

\subsection{List-mode data: ROOT files}
The Python side of GATE~10 manages ROOT files via the \verb|uproot| package~\citep{jim_pivarski_scikit-hepuproot_2020} when these are read as input, e.g., a phase space file for a source. Creating and writing ROOT files, on the other hand, is handled on the C++ side. GATE uses the Geant4 implementation of ROOT to avoid dependence on external C++ ROOT libraries and to ensure thread safety. A current limitation is that GATE~10 cannot store separate ROOT files per simulation run.  

For phase space sources, where primary particles are read from a ROOT file, particles are processed in a buffer (e.g., 10,000 particles at a time) on the Python side and then transferred to C++. Once the buffer is empty, the C++ side pauses and triggers the Python side to read more particles from the ROOT file. This approach is both efficient and convenient, especially in multithreaded environments where specific options allow different threads to start reading from different indices within the ROOT file.

\subsection{Regions, cuts, and step limiters}
\label{sec:regions_cuts}

Geant4 gives the user the possibility to fine-tune the particle tracking. In particular, the user can adjust cuts on the production of secondary particles (low energy secondaries are dumped locally) and configure the stepping mechanism, e.g. by imposing a maximum allowed step length. The older versions of GATE used their own \emph{ad hoc} mechanism to handle cuts and step limiters in nested volume trees. In GATE~10, we leveraged the Geant4 concept of Regions, i.e. logical groups of volumes sharing certain physics configurations. Regions are particularly useful in large geometries for which Geant4 originally developed that concept. 

GATE exposes advanced physics configurations, such as cuts and step limiters, to the user via simple commands and automatically handles the rather complex underlying Geant4 implementation.
Line 15 of code example~\ref{code:example-script} illustrates this, where the command \verb|patient.set_production_cut(particle_name="electron", value=3 * mm)| instructs Geant4 to produce and track only secondary electrons in volume \verb|patient| if their range is greater than 3\,mm. 
Internally, GATE~10 does the following: The object \verb|patient| contacts the \verb|PhysicsManager| and instructs it to create a production cut for electrons specific to volume ``patient''. The \verb|PhysicsManager| therefore creates a \verb|Region| object which handles physics parameters in the volume ``patient'' and sets the production cut of 3\,mm for electrons via the Region's user parameters. 
In this example, the Region handles a single volume only, but it can handle multiple volumes which share the same physics in a more complex simulation. 
When the simulation is executed via \verb|sim.run()|, the \verb|Region| object automatically creates the necessary Geant4 component (\verb|G4ProductionCuts|), which actually handles the production cuts and configures them (\verb|SetProductionCut()|). 

\subsection{Multithreading}
\label{sec:multithreading}

Monte Carlo simulations are intrinsically parallelizable because each primary particle is independent of the others. GATE~10 leverages this with multithreading (MT) and multiprocessing (MP) to accelerate simulations. We relied on Geant4's multithreading to share memory across threads while ensuring thread safety through strategies like thread locks and atomic operations. 
We implemented a modified, wrapped version of the Geant4 run manager to ensure that GATE~10 can correctly initialize and run simulations according to the Geant4 multithreading workflow. We also carefully handled Python's so-called global interpreter lock via the appropriate pybind11 configuration because we found that multithreaded simulations could otherwise lock up. 
Each actor explicitly ensures thread safety and flags itself as MT-compatible. 
Most GATE actors now support multithreading, though challenges persist in dynamic simulations and complex geometries due to geometry re-optimization. 

A significant technical challenge that we faced was that Geant4's multithreading mechanism collides with Python's global interpreter lock. GATE~10 runs multithreaded simulations by explicitly releasing this GIL which in turn was only possible thanks to the life cycle management described in section~\ref{sec:life-cycle-management}.


\subsection{Time-aware sources}


Geant4 keeps track of time for each particle, starting with the creation of the primary particle, but there is no temporal correlation among primary particles. On the other hand, 
the temporal structure of particle generation is critical in many medical physics applications, such as modeling physical and biological decay in radionuclide therapy or simulating systems that rely on time coincidences like PET, Compton cameras, or prompt-gamma timing systems. The time-handling engine in GATE 10 has been redesigned for greater flexibility. The simulation's macroscopic time is structured into one or more Geant4 runs defined by user-specified time intervals. For example, a user could define two distinct intervals: a first run from $t_0=10$~s to $t_1=30$~s (a duration of 20~s), and a second run from $t_2=50$~s to $t_3=60$~s (a duration of 10~s). GATE will only generate primary particles within these active time windows; no particles will be emitted in the gap between 30~s and 50~s. Additionally, sources in GATE are aware of the simulation time and assign the global time stamp to each generated particle. 

Each source is assigned an initial activity, for instance, 100~Bq at the start time $t_0$. In the context of a particle beam, this activity should be understood as the beam intensity. The particle generation proceeds according to this activity and the defined time structure of the simulation in terms of runs. If no decay is specified, the source activity remains constant. In the example above, the source would emit 100 particles per second. This results in expected values of $100 \times 20 = 2000$ particles in the first run and $100 \times 10 = 1000$ particles in the second run (the exact number of particles is stochastic). Each particle is assigned a timestamp corresponding to its generation time. 

When radioactive decay is enabled (e.g., for a $^{99\text{m}}$Tc source with a half-life of 6~hours), the activity is no longer constant. GATE updates the activity over time according to the law of radioactive decay: 
\begin{equation} 
A(t) = A_0 e^{-\lambda (t - t_0)} \label{eq:decay} 
\end{equation} 

where $A(t)$ is the activity at time $t$, $A_0$ is the initial activity at start time $t_0$, and $\lambda$ is the decay constant of the radionuclide. The emission of individual particles is a stochastic process. The time of the next particle emission, $t_{\text{next}}$, is sampled from an exponential distribution based on the current activity, $A(t_{\text{current}})$: 
\begin{equation} 
t_{\text{next}} = t_{\text{current}} - \frac{\ln(U)}{A(t_{\text{current}})} 
\label{eq:sampling} 
\end{equation}
where $t_{\text{current}}$ is the time of the last emission and $U$ is a random number drawn from a uniform distribution on the interval $(0, 1)$. This process ensures that particles are generated with a temporal distribution that accurately models the source's radioactive decay. 

If the primary particle is a Geant4 generic ion with a defined decay chain, its initial emission follows this timing mechanism, while any subsequent daughter products are emitted with their own physically accurate, stochastically determined timestamps which can be outside the user defined time intervals.

In addition to these models, users can modulate source strength using a Time Activity Curve (TAC). The TAC is defined as a discrete histogram, parameterized by vectors of time points and their corresponding activity values. The source's activity during the simulation is then determined by linearly interpolating between these points based on the current simulation time. This feature is useful for simulating dynamic processes in which the activity of a source changes over time in a complex manner.





%


\section{Discussion}

In this paper, we describe the code architecture and a series of methodological solutions that form the basis of the new Monte Carlo simulation software GATE~10. Many are related to the challenge of combining Python and C++, as well as to the complexity of the underlying Geant4 toolkit. 
The result is a software with which the user can set up and run particle and radiation transport simulations via a Python script and at the same time make use of the well-established capabilities of the Geant4 toolkit without time penalty. The software was released in November 2024 and had previously been available as a beta version for around 3 years. So far, users have reported that the Python interface is intuitive and efficient to use. 

As explained in section~\ref{sec:architecture}, GATE~10 contains a sub-package of wrapped C++ functions and classes, many of which come from Geant4, and it uses this wrapping internally to setup the Geant4 simulation. However, it would be highly impractical for the user to configure a simulation directly based on these wrappers. 
The user would also need to manually handle life cycle management of Python vs. C++ objects (see section~\ref{sec:life-cycle-management}) in each individual simulation. For similar reasons, packages such as \verb|g4py| and \verb|g4python| are unlikely to work without more extended code around them. In this sense, the architecture described in this article, together with GATE-specific functionality, such as sources and actors, is the crucial foundation of GATE~10 that turn it into much more than a Python-wrapper of Geant4 simulations. 

This environment makes it possible to implement complex systems related to medical physics as task-specific modules. For example, an institution could implement their therapeutic radiation source in a module that can then be imported by any script that simulates a patient treated with the particle source. An example is the IDEAL framework for proton therapy \citep{Grevillot2021}. External software, such as treatment planning systems or imaging software, can also import GATE~10 as a Python library. 
GATE~10's ability to retrieve output data directly from memory (and not only from disk) is a drastic enhancement compared to previous GATE versions and especially useful when embedding GATE~10 into other software.  

Although the development of GATE~10 represents an immense amount of work, numerous opportunities remain for further activities. One aspect concerns list-mode output, typically generated by digitizer actors,  which is currently stored as root files and cannot be made available directly in memory. We are considering alternative formats such as hdf5\footnote{\url{https://support.hdfgroup.org/documentation/hdf5/latest/index.html}} and will explore other mechanisms to ease the handling of list-mode data. 

Parallelization is a key to speeding up Monte Carlo simulations and GATE~10 currently implements multithreading to achieve this (see section~\ref{sec:multithreading}). Although multithreading is useful in many simulations, especially when executed on single machines, it also has some limitations. For example, when a simulation contains multiple Geant4 runs, Geant4's multithreading mechanism waits for a run to finish in all parallel threads until it progresses to the subsequent run. Consequently, the computation speed is limited by the slowest thread. Furthermore, multithreading currently only work on Unix-based platforms, but not on Windows, due to technical issues rooted in the operating system. 
An alternative way of parallelizing a simulation is to dispatch them to multiple processes at the same time, on a local computer in the simplest case, and to a larger computation infrastructure in a more complex scenario. This kind of parallelization can be achieved, e.g. by scripts, which launch multiple simulations and collect their output. To make this more convenient and efficient for the user, we are currently developing a mechanism in GATE~10 to automatically split and dispatch a simulation to multiple processes and recover and merge the output data afterward. The built-in serialization mechanisms (see section~\ref{sec:gateobject}) together with advanced handling of actor output (see section~\ref{sec:actoroutput}) will play a crucial role. 


The manager-engine structure makes it possible, in principle, to include alternative simulation back-ends to complement Geant4. This could be a task-specific fast tracking backend, e.g. specialized on proton dose calculation, which would be handled by a dedicated engine class. It would also be of interest to explore variance reduction techniques which could be triggered via the callback mechanism (see section~\ref{sec:callbacks}). For example, anytime a particle enters a particular volume, an effective generator function (possibly based on artificial intelligence) could be triggered to generate multiple realizations of the particle exiting the volume instead of explicitly tracking the particle across the volume. 

So far, the development of GATE~10 has focused on classic medical physics such as radiation therapy, medical imaging, and nuclear medicine. In principle, however, GATE~10 is able to simulate particle transport in any other context as long as the underlying Geant4 toolkit is capable of doing so. New applications likely require new ways of scoring physics properties, i.e. new actors. 
In the future, it may be possible to expand GATE~10's capability towards other areas as well, such as radiobiology and radiochemistry.

\section{Conclusion}
The newly developed version 10 of the Monte Carlo application GATE combines the robustness, speed, and versatility in terms of physics modeling of the Geant4 toolkit with a modern Python-based interface and advanced features. This synergy relies on a carefully crafted code architecture and dedicated solutions, which we have presented in this paper. We hope that GATE~10 will provide the community with a valuable modern tool to push their research toward new frontiers. 

The development has been the fruit of countless hours of work by a diverse group of researchers and engineers who have contributed their expertise, experience, and passion. The project's foundation as an open-source initiative is fundamental to this effort and fostering a collaborative spirit is and will be what drives the development of GATE forward. 

The release of GATE~10 in November 2024 has been a major milestone, but there are still many opportunities for future development and improvement. Ultimately, the development and maintenance of scientific software relies on funding, and we sincerely hope that this long-lasting collaborative project will also continue to be appreciated through financial support.

\section*{Acknowledgments}
The authors acknowledge funding support from the following grants: ANR-21-CE45-0026, INCa-INSERM-DGOS-12563, ANR-11-LABX-0063, ANR-11-IDEX-0007, and the Erasmus+ program. 

\clearpage

\bibliography{bibliography.bib}

\end{document}